\documentclass[12pt]{iopart}

\usepackage{graphicx}

\begin{document}

\title[Event anisotropy of identified $\pi^{0}$, photon and electron]{Event anisotropy of identified $\pi^{0}$, photon and electron
       compared to charged $\pi$, $K$, $p$ and deuteron in $\sqrt{s_{NN}}$ = 200 GeV Au+Au at PHENIX
      }

\author{Masashi Kaneta{\dag}
        for the PHENIX Collaboration\footnote[2]{For the full PHENIX Collaboration author list and acknowledgments, see Appendix ``Collaboration" of this volume.}
       }

\address{\dag\ RIKEN-BNL Research Center, Brookhaven National Laboratory, Upton NY 11973-5000, USA}

\ead{kaneta@bnl.gov}

\begin{abstract}
  We report the recent results of event anisotropy analysis focused on $v_2$ in $\sqrt{s_{NN}}$ = 200 GeV Au+Au collisions at PHENIX.
\end{abstract}



\section{Introduction}
  Event anisotropy analysis is a powerful tool for studying properties of the early stage in high-energy heavy ion collisions.
  Recent results of event anisotropy analysis suggest that finite $v_{2}$ (second harmonic coefficient of Fourier expansion of azimuthal distributions) is present up to relatively high transverse momentum ($p_{T}\sim3-4$ GeV/$c$)~\cite{nucl-ex/0307010}.
  Measurement of $\pi^{0}$'s can reach very high $p_T$ and thus should permit study of the effect of hard processes, e.g. jet quenching, on $v_2$.
  The dominant source of photons seen in the experiment is from meson decays (mainly from $\pi^{0}$'s), and there have been several measurements on $v_2$ of charged hadron and identified $\pi^{0}$ up to high $p_T$ close to 10 GeV/$c$.
  As a next step in the systematic study of $v_2$ at high $p_T$, we have investigated subtracting the hadronic decay contribution from the $v_2$ of the measured photons to determine the feasibility of seeing the direct or thermal photon $v_2$.
  We have also investigated the measured electron spectrum, which includes contributions from hadron decays and conversion photons (referred to as ``\textit{photonic electrons}").
  These effects are well studied~\cite{PRL_88_2002_192303,Sean_Kelly} and we extract the charmed electron distribution after subtractions.
  We adopt the same method in electron $v_2$ analysis to obtain the charmed electron $v_2$.

\section{Analysis}
  Event anisotropy analysis is an application of the Fourier expansion of azimuthal particle distributions.
  Here we focus on the elliptic component of the event anisotropy.
  Due to momentum conservation and rotational symmetry, the odd order harmonics will be canceled when we add positive and negative rapidity regions.
  Additionally assuming that the contribution of higher order of harmonics is negligible, the distributions are described by
  \begin{equation}
  \fl
   E \frac{dN^{3}}{d^{3}p}  =   \frac{1}{2{\pi}}\frac{d^{2}N}{p_{T} dp_{T} dy}
                                  \Bigl(
                                    1 + 2 v_{2}^{m} \cos[2(\phi-\Phi_{r})]
                                  \Bigr)
                                  \nonumber 
                           =    C
                                  \Bigl(
                                    1 + 2 v_{2}^{m} \cos[2(\phi-\Phi_{r})]
                                  \Bigr),
   \label{eq_fourier}
  \end{equation}
  where $y$ is the rapidity and $\phi$ is the azimuthal angle of the particle, $\Phi_{r}$ is the reaction plane, $v_{2}^{m}$ is the 2nd order of the Fourier coefficient measured by detectors, and $C$ is a constant.
  The reaction plane is defined using Beam-Beam counters~\cite{NIM_A411_1998_238} ($|\eta|$=3.1$-$4.0) with a resolution due to signal size of $v_2$ and multiplicity on the detector.
  Therefore, $v_{2}^{m}$ is broadened by the uncertainty of the reaction plane definition.
  The method of $v_{2}^{m}$ correction we adopted, described in reference~\cite{PRC58_1998_1571}, results in a real $v_{2}$ corrected by the reaction plane resolution $v_{2} = v_{2}^{m} / \sigma_{2}$, where $\sigma_{2}$ is the resolution for the 2nd harmonic.
  Since the reaction plane is distributed uniformly, the efficiency of the detector is canceled.
  Hence, we fit the number of particles as a function of $\phi-\Phi_{r}$ by equation~(\ref{eq_fourier}) to obtain $ v_{2}^{m}$.

  The $\pi^{0}$ $v_{2}$ analysis is performed in the following steps:
  1) counting the number of $\pi^{0}$'s from invariant mass from two photons measured by the EMCal~\cite{hep-ex/0304038,PRL91_2003_072301} as a function of $\phi-\Phi_{r}$ for each centrality and $p_{T}$ bin,
  2) fitting the distribution by equation~(\ref{eq_fourier}), and
  3) reaction plane resolution correction to $v_{2}^{m}$.
  The photon and electron $v_{2}$ analysis are checked by comparison to another method, which is $v_{2}^{m} = \langle\cos[2(\phi-\Phi_{r})]\rangle$.
  The difference between two methods is taken into account as a systematic error.
  The additional contributions to the systematic error are the following:
  (1) Particle identification cut,
  (2) reaction plane determination, and
  (3) error propagation in ``\textit{photonic electron}" subtraction (only for charmed electron $v_2$).
  The total systematic error is estimated by adding the contributions quadratically.

\section{Results and Discussion}

\begin{figure}[t]
    \begin{center}
      \includegraphics[scale=0.54]{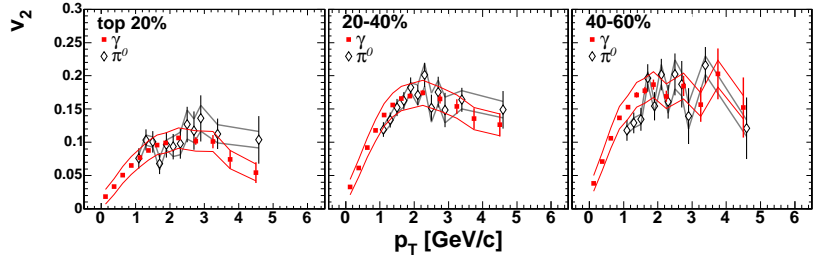}
    \end{center}
    \caption{\label{photon_and_pi0_v2_vs_pt}
             (Color online)
             Preliminary results for photon (solid square) and $\pi^{0}$ (open diamond) $v_2$ as a function of $p_T$ from three centrality bins in $\sqrt{s_{NN}}$ = 200 GeV Au+Au collisions.
             The statistical error is shown as a vertical bar on each point.
             Systematic errors are shown by curves.
            }
  \end{figure}

  Figure~\ref{photon_and_pi0_v2_vs_pt} shows photon and $\pi^{0}$ $v_{2}$ as a function of $p_{T}$ from three centrality bins (top 20\%, 20-40\%, and 40-60\%) in $\sqrt{s_{NN}}$ = 200 GeV Au+Au collisions.
  The upper limit of the $p_{T}$ range is defined by the available statistics while
  the lower limit of $p_{T}$$>$1 GeV/$c$ results from hadron contamination in the EMCal.
  The $\pi^{0}$ $v_{2}$ increases with $p_{T}$ and then saturates at $p_{T}$ $\sim$ 3 GeV/$c$ in each centrality.
  Comparison to the PHENIX data for identified charged pion $v_2$~\cite{nucl-ex/0305013}
  indicates good consistency with the $\pi^{0}$ $v_2$ results presented here (shown in reference~\cite{MK_DNP2003}).
  Within errors, the photon $v_2$ is consistent with that of the $\pi^{0}$ $v_2$.
  Better statistics is necessary to obtain the direct photon $v_2$, especially at high $p_T$.

\begin{figure}[t]
    \begin{center}
      \includegraphics[scale=0.37]{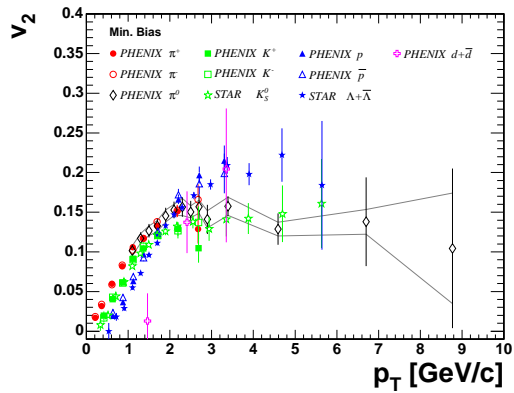}
    \end{center}
    \caption{\label{v2_vs_pt_mb_summary}
             (Color online)
             Preliminary results for $\pi^{0}$ $v_2$ as a function of $p_T$ from minimum bias data 
             in $\sqrt{s_{NN}}$ = 200 GeV Au+Au collisions, compared to other  hadrons.
             The statistical error is shown as a vertical bar on each point and the systematic error is shown by curves for the $\pi^{0}$'s.
             The other hadrons are shown with statistical error (a vertical bar) only.
             The charged $\pi$, $K$, and $p$ $v_2$ are from reference~\cite{nucl-ex/0307010}.
             The deuteron $v_2$ is PHENIX preliminary data shown in reference~\cite{SSakai_HIC03}.
             The $K_{S}^{0}$ and $\Lambda$ $v_2$ are from STAR in reference~\cite{PRL_92_2004_052302}.
            }
  \end{figure}

  The result for $\pi^{0}$ $v_2$ shown in Figure~\ref{v2_vs_pt_mb_summary} is the first measurement of an identified hadron 
up to $p_{T}$ = 10 GeV/$c$ in minimum bias 200 GeV Au+Au collisions.
  It has a non-zero value up to $p_{T}$ = 8 GeV/$c$.
  The comparison is done with other hadrons from results of PHENIX~\cite{nucl-ex/0307010,SSakai_HIC03} and the STAR collaboration~\cite{PRL_92_2004_052302}.
  There is a crossover point among hadrons around $p_T$ = 2 GeV/$c$.
  The lower $p_T$ ($<$2 GeV/$c$) region is well described by hydrodynamical models~\cite{nucl-ex/0307010,PRL_92_2004_052302}.
  In the region of $p_T$$>$2GeV/$c$, the pion $v_2$ merges with that of the kaon, as does the proton and $\Lambda$ $v_2$.
  At the same time, the  meson $v_2$ remains different from that for baryons.
  This suggests a scaling rule called quark coalescence~\cite{quark_coalescence}, related to the different number of quarks in mesons and baryons, which might indicate that the hadron $v_2$ is generated during the early partonic stage before the hadronization.
  A simple check of a coalescence picture in $v_2$ is normalization of both $v_2$ and $p_T$ by the number of constituent quarks.


\begin{figure}[t]
    \begin{center}
      \includegraphics[scale=0.325]{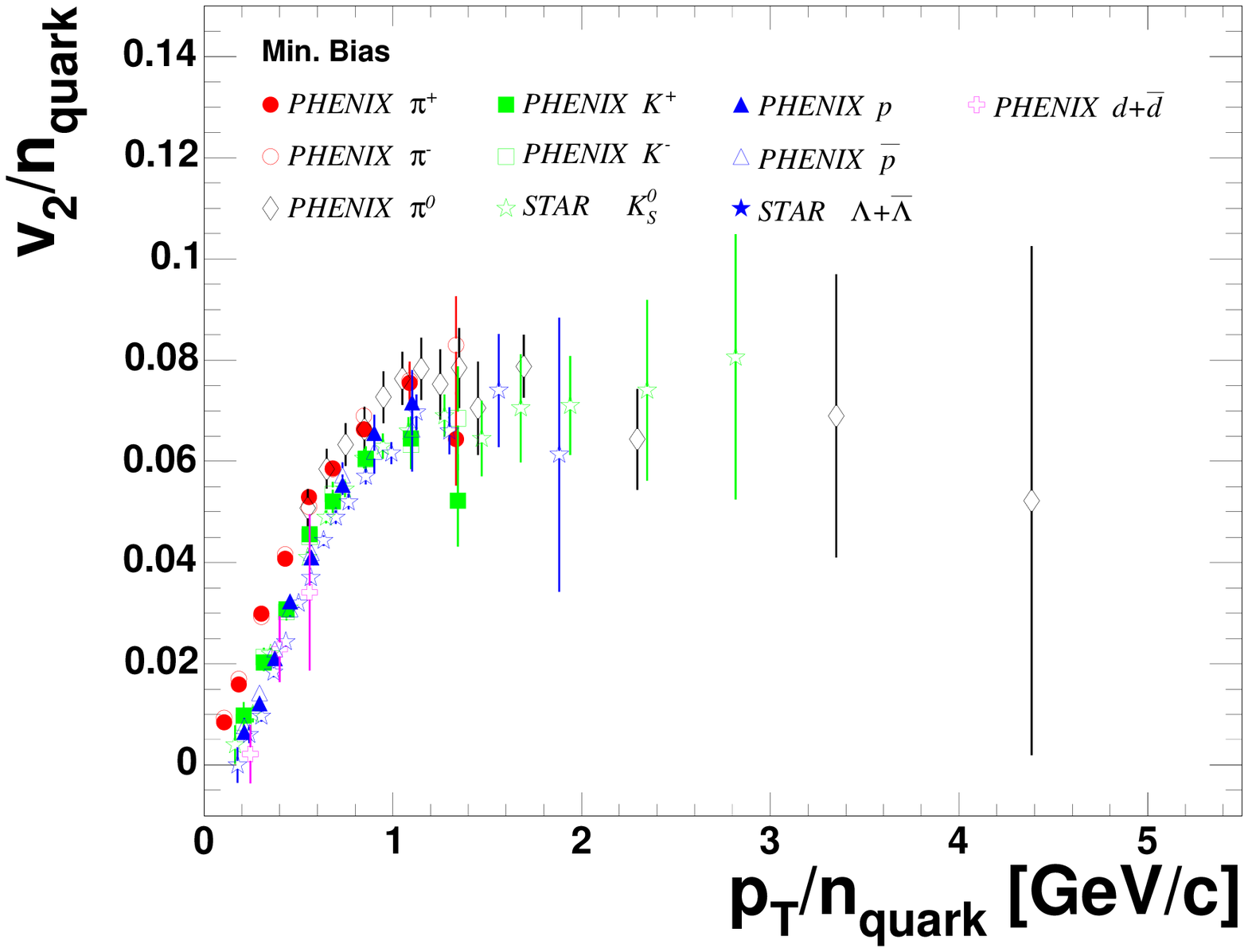}
      \includegraphics[scale=0.355]{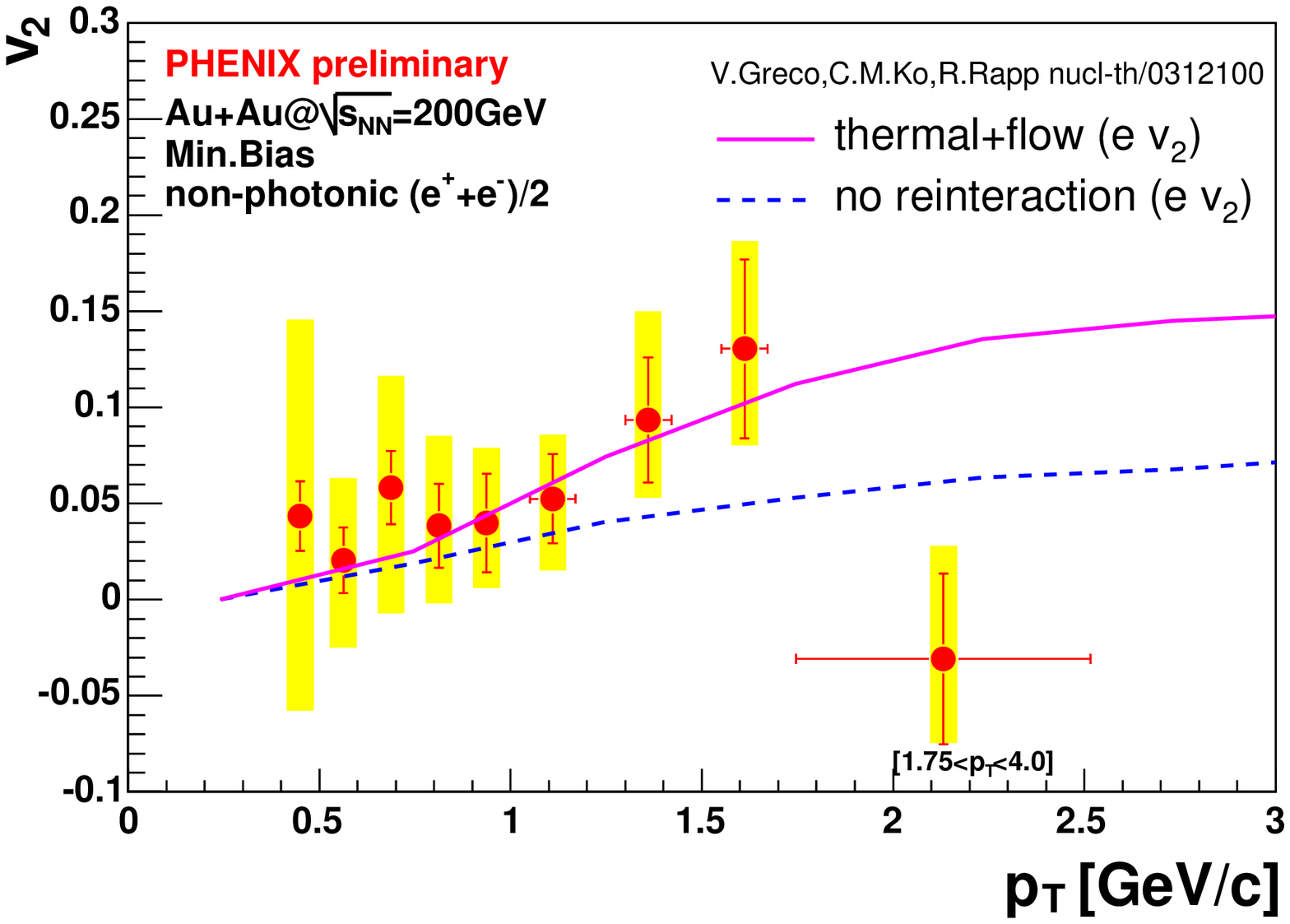}
    \end{center}
    \caption{\label{v2n_vs_ptn_mb_summary}
             (Color online)
             The left-hand side plot shows $v_2$ as a function of $p_T$ scaled via the coalescence prescription.
             The meson, Baryon, and deuteron are scaled by 2, 3, and 6, respectively.
             ``Non-photonic" electron $v_2$ as a function of $p_T$ is shown at the right-hand side.
             The vertical bar on each point show the statistical error and systematic errors are shown by colored box.
             The horizontal bar shows the RMS of $dN/dp_T$ in each bin. 
            }
  \end{figure}

  The left-hand side plot in Figure~\ref{v2n_vs_ptn_mb_summary} is the plot to check the coalescence picture in $v_2$.
  The data from $\pi$, $K$, $p$, $\Lambda$ are superimposed and show good agreement in $p_T/n_{quark}>1$ GeV/$c$.
  The plot shows that the deuteron $v_2$ agrees with the other hadrons by scaling number of quarks.  
  However, we note that nucleon coalescence could be the dominant factor for forming deuterons rather than the quark coalescence; there is no way to separate them from deuteron $v_2$ measurements alone.

  The ``non-photonic" electron $v_2$ vs. $p_T$ in minimum bias $\sqrt{s_{NN}}$ = 200 GeV collisions is shown on the right-hand side of Figure~\ref{v2n_vs_ptn_mb_summary} with two scenarios from a model calculation from reference~\cite{nucl-ex/0312100}.
  One scenario is that $D$ mesons are made from completely thermalized charm and light quarks (shown by solid line), while the other is that there is no interaction of the $c$-quarks, so that the flow contribution in $D$'s is only from the light quarks (dashed line) .
  The non-photonic electrons decay from charmed and bottomed mesons; those in the $p_T$ range we showed are dominantly charmed meson decay.
  (A detailed discussion of the electron measurement is found in reference~\cite{Sean_Kelly}.)
  Within the available statistics, both models are consistent with data.
  It will be very interesting to pursue this comparison of charm flow with that of mesons containing light quarks using the higher statistics Run-4 data set.


\section*{References}
  \begin{thebibliography}{10}
   \bibitem{nucl-ex/0307010}
     PHENIX Collaboration K. Adcox \etal 2003 \textit{Preprint} nucl-ex/0307010
   \bibitem{PRL_88_2002_192303}
     PHENIX Collaboration K. Adcox \etal 2002 \PRL \textbf{88} 192303
   \bibitem{Sean_Kelly}
     PHENIX Collaboration S. Kelly \etal in this proceedings
   \bibitem{NIM_A411_1998_238}
      K.~Ikematsu \etal 1998 Nucl. Instrum. Meth. A \textbf{411} 238
   \bibitem{PRC58_1998_1571}
     A.M. Poskanzer and S.A. Voloshin 1998 \PR C\textbf{58} 1671
   \bibitem{hep-ex/0304038}
     PHENIX Collaboration S.S. Adler \etal 2003 \textit{Preprint} hepl-ex/0304038
   \bibitem{PRL91_2003_072301}
     PHENIX Collaboration S.S. Adler \etal 2003 \PRL \textbf{91} 072301
   \bibitem{nucl-ex/0305013}
     PHENIX Collaboration S.S. Adler \etal 2003 \textit{Preprint} nucl-ex/0305013
   \bibitem{MK_DNP2003}
     PHENIX Collaboration M. Kaneta \etal 2003 DNP2003 talk.
   \bibitem{SSakai_HIC03}
     PHENIX Collaboration S. Sakai \etal 2003 HIS03 poster.
   \bibitem{PRL_92_2004_052302}
     STAR Collaboration J. Adams \etal 2004 \PRL \textbf{92} 052302
   \bibitem{quark_coalescence}
     For example, V. Greco, C.M. Ko and P. Levai 2003 \PR C \textbf{68} 034904 : 
     R.J.~Fries, B.~Muller, C.~Nonaka and S.A.~Bass 2003 \PRL \textbf{90} 202303 :
     D. Moln\'ar and S.A. Voloshin 2003 \PRL \textbf{91} 092301
   \bibitem{nucl-ex/0312100}
     V. Greco, C.M. Ko and R. Rapp 2003 \textit{Preprint} nucl-th/0312100
  \end{thebibliography}
\end{document}